\documentclass[aps,pre,onecolumn,superscriptaddress,amsmath,amssymb,letterpaper,10pt]{revtex4-1}
\usepackage[utf8]{inputenc}                
\usepackage[T1]{fontenc}
\usepackage{graphicx}
\usepackage{epstopdf}
\usepackage{bm}
\usepackage{pdfpages}
\begin{document}
\title{A shadowing problem in the detection of overlapping  communities: lifting the resolution limit  through a cascading procedure}
\author{Jean-Gabriel Young}
\affiliation{D\'epartement de Physique, de G\'enie Physique, et d'Optique, Universit\'e Laval, Qu\'ebec (Qu{\'e}bec), Canada, G1V 0A6}
\author{Laurent H\'ebert-Dufresne}
\affiliation{D\'epartement de Physique, de G\'enie Physique, et d'Optique, Universit\'e Laval, Qu\'ebec (Qu{\'e}bec), Canada, G1V 0A6}
\affiliation{Santa Fe Institute, Santa Fe, NM 87501, USA}
\author{Antoine Allard}
\affiliation{D\'epartement de Physique, de G\'enie Physique, et d'Optique, Universit\'e Laval, Qu\'ebec (Qu{\'e}bec), Canada, G1V 0A6}
\affiliation{Departament de F\'eisica Fonamental, Universitat de Barcelona, Barcelona, Espanya, 08028}
\author{Louis J. Dub\'e}
\affiliation{D\'epartement de Physique, de G\'enie Physique, et d'Optique, Universit\'e Laval, Qu\'ebec (Qu{\'e}bec), Canada, G1V 0A6}

\begin{abstract}
Community detection is the process of assigning nodes and links in significant communities (e.g. clusters, function modules) and its development has led to a better understanding of complex networks.
When applied to sizable networks, we argue that most detection algorithms correctly identify prominent communities, but fail to do so across multiple scales.
As a result, a significant fraction of the network is left uncharted.
We show that this problem stems from larger or denser communities overshadowing smaller or sparser ones, and that this effect accounts for most of the undetected communities and unassigned links.
We propose a generic cascading approach to community detection that circumvents the problem.
Using real and artificial network datasets with three widely used community detection algorithms, we show how a simple cascading procedure allows for the detection of the missing communities.
This work highlights a new detection limit of community structure, and we hope that our approach can inspire better community detection algorithms.
\end{abstract}

\maketitle

\section{Introduction}
Over the course of the last decade, network science has attracted an ever growing interest since it provides important insights on a large class of interacting complex systems.
One of the features that has drawn much attention is the structure of interactions highlighted by the network representation.
Indeed, it has become increasingly clear that global structural patterns emerge in most real networks \cite{Girvan2002}.
One such pattern, where links and nodes are aggregated into larger groups, is called the community structure of a network.

While the exact definition of communities is still not agreed upon \cite{Fortunato2010}, the general consensus is that these groups should be denser than the rest of the network.
The notion that communities form some sort of independent units (families, circles of friends, coworkers, protein complexes, etc.) within the network is thus embedded in that broader definition.
It follows that communities represent functional modules, and that understanding  their layout as well as their organization on a global level is crucial to a fuller understanding of the system under scrutiny \cite{Hebert2010, Hebert2011}.

By developing techniques to extract this organization, one assumes that communities are encoded in the way nodes are interconnected, and that their structure may be recovered from limited and/or incomplete topological information.
Various algorithms and models have been proposed to tackle the problem, each featuring a different definition of the community structure while sharing the same general objective.
Although these tools have been used with success in several different contexts \cite{Fortunato2010,Ahn2010,Palla2005}, a number of shortcomings are still to be addressed.

The paper is organized as follows.
First, we argue that current algorithms tend to overlook small communities found in the neighborhood of larger, denser ones, under very general conditions.
In the following sections, we investigate the exact mechanisms that cause this so-called \emph{shadowing} phenomenon in 3 widely used algorithms.
Then, we propose and develop a general \emph{cascading} approach to community detection that addresses this specific problem. 
Next, we validate our first implementation of the cascading method by applying our algorithms to real complex networks, and then to synthetic benchmark networks.
The former set of calculations acts as a proof of concept and shows that, despite its apparent simplicity,  the algorithm detects several shadowed communities. 
The latter set helps us determine under which conditions the algorithm is expected to perform adequately.
Finally, based on our results, we gather in the Conclusion ways in which our method can be improved.
%
\section{Resolution limit due to shadowing}
%
It is known that a resolution limit exists for a large class of community detection algorithms that rely on the optimization of a quality function over \emph{non-overlapping} partitions of the network.
This resolution limit has been rigorously shown to affect modularity \cite{Newman2004,Fortunato2007}, map equation \cite{Rosvall2008,Kawamoto2015} and description length \cite{Peixoto2013} based methods.
It stems from the fact that the number or the size of the smallest detectable community, is related to the size of the network \cite{Fortunato2007,Peixoto2013,Traag2011}.
As a result, clearly separated clusters of nodes are sometimes considered as one larger community, because they are too small to be resolved by the detecting algorithm.
A suggested solution \cite{Fortunato2007} is to conduct a second analysis on all detected communities to verify that no smaller internal modules can be identified.

In the majority of real world applications, the optimal covering of a network should include \emph{overlapping communities}, since they capture the multiplicity of functions that a node might fulfill \cite{Ahn2010}.
We argue that in the overlapping case, a different resolution limit arises in detection algorithms that \textit{rely on some global resolution parameter}, due to an effect that we refer to as \textit{shadowing}.
The resolution parameter may be implicit or explicit, fixed or flexible.

In essence, shadowing occurs when large/dense communities act as screens preventing the detection of smaller/sparser adjacent communities.
To illustrate this phenomenon, we study three detection algorithms based on two different paradigms of community structure, namely nodes and links communities.
Note that while improved versions of these algorithms have been proposed \cite{Evans2009,Farkas2007,Kumpula2008,Xie2013}, none raises, let alone addresses the shadowing problem.

\subsection{Clique percolation algorithm} 
The clique percolation algorithm (CPA) \cite{Palla2005} defines communities as  maximal \textit{$k$-clique} percolation chains, where a $k$-clique is a fully connected subgraphs of $k$ nodes, and where a percolation chain is a group of cliques that can be reached from one adjacent $k$-clique to another \cite{Derenyi2005}.
Two $k$-cliques are said to be adjacent if they share $k-1$ nodes.
The complete community structure is obtained by detecting every maximal percolation chains for a given value of $k$.
Because percolation chains consist of $k$-cliques sharing $k-1$ nodes, overlapping communities occur whenever two cliques share less than $k-1$ nodes.
It is noteworthy that the definition of a community in this context is consistent with the general description of communities outlined in the Introduction.
Indeed, $k$-clique percolation chains are dense by definition, and a sparser neighboring region is required to stop a $k$-clique percolation chain, ensuring that communities are denser than their surroundings.

\begin{figure}
    \includegraphics[width=0.8\linewidth]{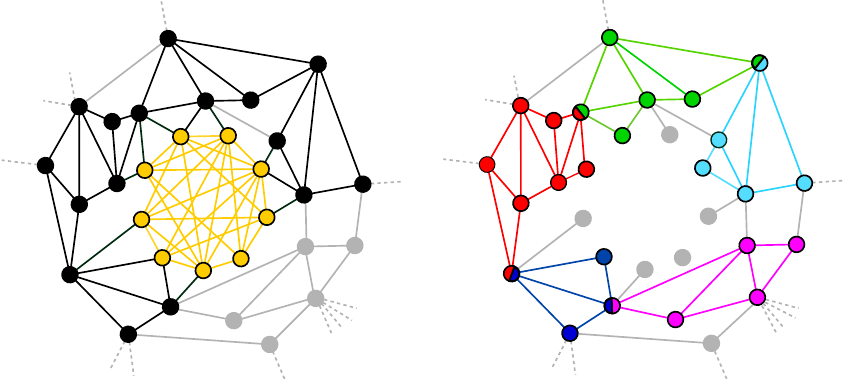}
  \caption{{\bf Shadowing effect for the CPA.}
  \emph{Left panel:} The yellow region is the sole detectable community with $k=4,5$, while its union with the black region corresponds to the community detected with $k=3$.
  This pathological example illustrates the two undesirable extreme effects mentioned in the main text: either most of the network is detected as a single community, or only large and dense clusters are detected.
  No optimal value of $k$ can be found in this case.
  \emph{Right panel:} The structure of this subgraph nevertheless suggests that it could be decomposed in a dense community in the middle, surrounded by smaller communities.
  If the links involved in the dense community (detected with $k=4$ or $5$) were removed, a second iteration of the algorithm  with $k=3$ would lead to the detection of several smaller communities that were overshadowed by the larger one.}
  \label{Fig1}
\end{figure}

We expect shadowing since the size of the cliques, $k$, acts as a global resolution parameter.
Indeed, in principle, low values of $k$ lead to a more flexible detection of communities as a smaller clique size allows a wider range of configurations.
However, low values of $k$ often yield an excessively coarse-grained community structure of the network since percolation chains may grow almost unhindered and include a significant fraction of the nodes.
In contrast, large values of $k$ may leave most of the network uncharted as only large and dense clusters of nodes are then detected as communities.
An \textit{optimal value} corresponding to a compromise between these two extreme outcomes must therefore be chosen.
For the purpose of this study, we use the lowest value of $k$ such that no extensive community is detected.
As suggested in \cite{Palla2005}, the largest community is considered extensive if it contains about twice as many nodes as the second largest community.
As this value of $k$ attempts to balance two unwanted effects for the network as a whole, a shadowing effect is expected to arise causing the algorithm to overlook smaller communities, or to merge them with larger ones.
See Fig \ref{Fig1} for an illustration of this effect.

\subsection{Greedy clique expansion} 

The Greedy Clique Expansion (GCE) algorithm \cite{Lee2010} is superficially similar to the CPA.
Both algorithms grow communities starting from a set of very dense initial communities, or seeds.
However, GCE relaxes the definition of communities used in CPA in two different ways.
These slight modifications propel GCE to the status of a ``state of the art'' algorithm that works well even for hard detection problems \cite{Xie2013}.
First, GCE uses maximal cliques of at least $k$ nodes as seeds whereas CPA is seeded with cliques of a fixed number of nodes, which therefore does not exclude embedded cliques.
Second, GCE uses a less stringent growth process: it maximizes a fitness function instead of proceeding through strict clique percolation chains.
Essentially, GCE expands maximal cliques by sequentially incorporating the nodes that increase the fitness of a community \cite{Lancichinetti2009b}.
It follows that communities naturally overlap, because they are expanded locally, independent of each other.
In fact, the amount of overlap is so important that one must discard the communities that are too similar \cite{Lee2010}.

\begin{figure}
    \includegraphics[width=0.5\linewidth]{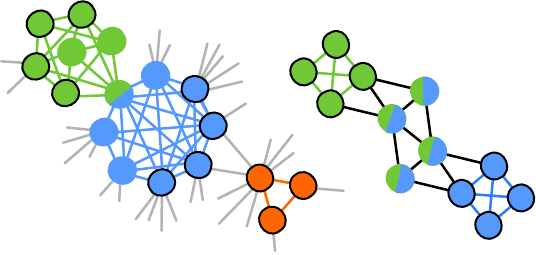}
  \caption{
  {\bf Shadowing effect in GCE.}
  \emph{Left panel:} The seeds (circled nodes) are expanded into two slightly overlapping communities (blue and green nodes) with $k=4$.
  The orange triangle is not merged with the blue community, nor assigned, unless one select seeds of size $k\geq3$, which is not always possible if most of the network is dense. 
  In such case. this value of $k$ would lead to the detection of large, highly redundant and meaningless communities.
  \emph{Right panel:} GCE expands the green and blue seeds (circled) to include the four nodes in the middle.
  The two communities are too similar, and one is discarded.   This leaves the other maximal clique unassigned.
  }
  \label{Fig2}
\end{figure}

The pervasive overlap and the fact that $k$ acts as a resolution parameter \emph{both} lead to shadowing.
Indeed, to avoid excessively large and redundant communities, one must typically use maximal cliques of size $k\geq 4$ as starting seeds.
Isolated triangle based communities that cannot be reached by expanding a nearby seed are therefore overlooked.
More interestingly, maximal-cliques are sometimes discarded, since the algorithm ignores strongly overlapping communities.
Both of these effects are illustrated in Fig \ref{Fig2}.

%
\subsection{Link clustering algorithm}
\begin{figure}
    \includegraphics[width=0.35\linewidth]{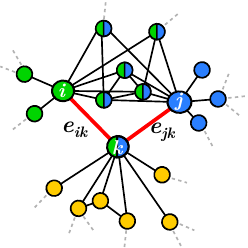}
  \caption{{\bf Calculation of the similarity between two links.}
  The sets $n_+\left(i\right)$ and $n_+\left(j\right)$ are respectively colored in green and blue.
  From Eq.~\eqref{Eq:Similarity_Ahn}, we have $S(e_{ik},e_{jk})=6/13$.
  Note that apart from nodes $i$ and $j$, the neighboring nodes of the keystone $k$ (colored in yellow) are not considered in the calculation of $S(e_{ik},e_{jk})$.}
  \label{Fig3}
\end{figure}

%
The link clustering algorithm (LCA) \cite{Ahn2010} aggregates links --- and hence the nodes they connect --- into communities based on the similarity of their respective neighborhood.
Denoting $e_{ab}$ the link between nodes $a$ and $b$, the similarity of two adjacent links $e_{ik}$ and $e_{jk}$ (attached to a same node $k$ called the \textit{keystone}) is quantified through a Jaccard index
\begin{equation} \label{Eq:Similarity_Ahn}
  S\left(e_{ik},e_{jk}\right) = \frac{\left|n_+(i) \bigcap n_+(j)\right|}{\left|n_+(i) \bigcup n_+(j)\right|} \ ,
\end{equation}
where $n_+(q)$ is the set of node $q$ and its neighbors, and $|n_+(q)|$ is the cardinality of the set.
Fig \ref{Fig3} illustrates the calculation of $S\left(e_{ik},e_{jk}\right)$.

Once the similarity has been calculated for all adjacent pairs of links, communities are built by iteratively aggregating adjacent links whose similarity exceeds a given threshold $S_c$.
This algorithm naturally allows communities to overlap (to share nodes) since a node can belong to as many communities as its degree.

Again, a shadowing effect is expected, since the similarity threshold $S_c$ acts as a global resolution parameter.
To elucidate the global aspect of $S_c$, one must describe how its value is chosen (as proposed in \cite{Ahn2010}).
Let us first define the density $\rho_j$ of community $j$ as
\begin{equation} \label{Eq:Community_Density}
  \rho_j=\frac{d_j-(n_j-1)}{\frac{n_j(n_j-1)}{2}-(n_j-1)} \ ,
\end{equation}
where $d_j$ and $n_j$ are the number of links and nodes in community $j$, respectively.
Considering that a community of $n$ nodes must at least include $n-1$ links, $\rho_j$ computes the fraction of potential ``excess links'' that are present in the community.
The similarity threshold $S_c$ is chosen to maximize the overall density of the communities
\begin{equation} \label{Eq:Partition_Density}
 \rho(S_c)=\frac{1}{D}\sum_{j\in\mathcal{C}(S_c)} d_j \rho_j
\end{equation}
where $\mathcal{C}(S_c)$ is the set of communities detected for a given $S_c$, and where $D$ is the total number of assigned links.
$\rho(S_c)$ is typically a well-behaved function of $S_c$ that displays a single maximal plateau \cite{Ahn2010}.
The value of $S_c$ corresponding to this plateau is selected since it leads, \emph{on average}, to the denser set of communities, hence its global nature.

\begin{figure}
    \includegraphics[width=0.3\linewidth]{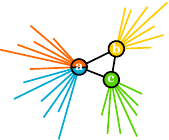}
  \caption{{\bf Shadowing effect for the LCA.}
  The pairwise unions of the three sets $n_+(a)$, $n_+(b)$ and $n_+(c)$ contain considerably more elements that their corresponding intersections since nodes $a$, $b$ and $c$ all have high degrees.
  According to Eq.~\eqref{Eq:Similarity_Ahn}, this implies that $e_{ab}$, $e_{bc}$ and $e_{ac}$ share lower similarities --- namely $S(e_{ac},e_{bc})=S(e_{ab},e_{bc})=3/22$ and $S(e_{ab},e_{ac})=3/17$ --- than if the triangle had been completely isolated ($S(e_{ac},e_{bc})=S(e_{ab},e_{bc})=S(e_{ab},e_{ac})=1$).
  It is therefore likely that these three links will be left unassigned.}
  \label{Fig4}
\end{figure}

Following an analysis similar to that presented in the CPA case, we expect small communities to be left undetected as they are eclipsed by larger and denser ones.
This is mainly due to the use of a resolution parameter ($S_c$) that cannot be adjusted locally.
For instance, links in a small community could exhibit vanishing similarities because some of the associated nodes are hubs (nodes of high degree).
This is especially true in the vicinity of large and dense clusters whose nodes are typically of high degree (see Fig \ref{Fig4} for an illustration).

%
%
\section{Cascading detection}
%
Figs \ref{Fig1} and \ref{Fig4} suggest that the inability to detect small or sparse communities in the vicinity of larger or denser ones (the shadowing effect) could be circumvented by removing the obstructing structures from the networks.
We formalize this idea and propose a \textit{cascading} approach to community detection that proceeds as follows:
\begin{enumerate}
 \item identify large or dense communities by tuning the resolution parameter;
 \item remove the internal links of the communities identified in step 1;
 \item repeat until no new significant communities are found.
\end{enumerate}
The first iteration of this algorithm detects the communities that are normally targeted by detection algorithms, thus ensuring that the cascading approach retains the main features of the ``canonical'' community structure.
After removal of links involved in the detected communities, a new iteration of the detection algorithm is then performed on a sparser network in which previously hidden communities are now apparent.
This process is repeated until a final and more thorough covering of the network into overlapping communities is obtained.
Note that the resulting cover is not necessarily hierarchical, but simply more complete.

For example, in the case of the CPA, a high value of $k$ (which leads to the traditional community structure) is selected for the first iteration of the algorithm.
The network then becomes significantly sparser since all cliques of size $k'>k$ are destroyed by the removal of internal links in step 2.
Subsequent iterations of the detection algorithm can then be conducted at lower $k$, unveiling finer structures, as the pathways formed by dense cluster are no longer available.
The process naturally comes to a halt at $k=3$, since $k=2$ only detects the disjoint components of the network.

A similar strategy is employed to uncover hidden communities with GCE.
Seeds of at least $k=4$ nodes are first used, until no seeds remain.
Then, seeds of smaller size $k=3$ are used.
Once again, the process halts when no seeds remain.
In the case of the LCA, the detection is stopped \emph{before} the partition density reaches zero, because $\rho(S_c)\simeq0$ only yields chains of links (the keystone ensures a non-vanishing similarity), which in general are not classified as significant communities.

It is worth mentioning that conducting this repeated analysis does not increase the computational cost significantly, because the cascading algorithm scales exactly like the community detection algorithm used at each iteration, and because the number of iterations that can be carried is small (typically less than 10).
Moreover, the size of the networks (number of links and nodes) effectively decreases after each iteration, further reducing the cost (numerical evidences will be presented in the next Section).
%
%
%
%
%
\section{Results and discussions}
%

\subsection{Real networks}
%
To investigate the efficiency and the behavior of the cascading detection, we first apply our approach to 8 small real network datasets: arXiv cond-mat circa 2004 (hereafter: \emph{arXiv}) \cite{Palla2005}, University Rovira i Virgili email exchanges (\emph{Email}) \cite{Guimera2003}, Gnutella peer-to-peer data (\emph{Gnutella}) \cite{Ripeanu2002}, internet autonomous systems (\emph{Internet}) \cite{Hebert2011}, Pretty-Good-Privacy data exchange (\emph{PGP}) \cite{Boguna2004}, Western States Power Grid (\emph{Power}) \cite{Watts1998}, Protein-protein interactions (\emph{Protein}) \cite{Palla2005} and word associations (\emph{Words}) \cite{Palla2005}.
Their properties are summarized in the Supporting Information (See S1 Table).

\begin{figure}
    \includegraphics[width=0.8\linewidth]{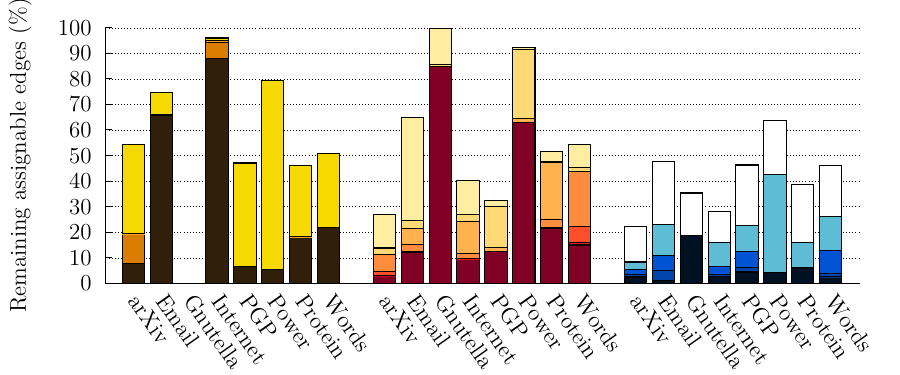}
    \caption{{\bf Fraction of remaining assignable links for real networks using the cascading approach.}
    (\textit{Left}) The number of unassigned links after one iteration of the CPA (corresponding to a typical use) is shown in yellow, and the final state is shown in dark brown.
    Whenever more than 2 iterations were performed, the intermediate results are shown in orange.
    For the \textit{Gnutella} network, the optimal value was $k=3$ at the first iteration, leading to an immediate complete detection of the community structure.
    For the purpose of selecting $k$, we consider that a cover contains an extensive community if the largest community is twice as large as the second largest community.
    In the case of the \textit{Internet} and \textit{Protein} networks, which contains large unbreakable clique, we used a looser criterion ($c \cdot n_{\text{largest}} < n_{\text{2nd largest}}$, with $c = 0.25$ and $c=0.30$, respectively).
    (\textit{Center}) Results of a canonical use of GCE are shown in beige and shades of red correspond to subsequent iterations.
    The final state is shown in dark red.
    (\textit{Right}) Results of a canonical use of the LCA are shown in white and shades of blue correspond to subsequent iterations.
    The final state is shown in dark blue.
    Note that all results are normalized to the number of assignable links in the original network.
    For the CPA, this corresponds to the number of links that belong to at least one 3-clique.
    For GCE, this corresponds to the number of links that belong to a component that contains at least one $k$-clique $(k\geq3)$.
    For the LCA, a link is considered assignable if at least one of the two nodes it joins have a degree greater than one.
    Numerical results are summarized in the Supporting Information (S2 Table).}
    \label{Fig5}
\end{figure}

First and foremost, our results show that cascading detection \emph{always} improves the thoroughness of the community structure detection.
Fig \ref{Fig5} shows that while a traditional use of the algorithms yields partitions with high fractions of unassigned links, the cascading approach leads to community structures where this fraction is significantly reduced.
More precisely, the percentage of remaining assignable links drops from 54.1\% to 26.3\% on average in the case of CPA, from 57.8\% to 27.7\% in the case of GCE, and from 41.0\% to 5.3\% in the case of LCA.
Note how cascading detection is more efficient when applied to the LCA.
This is due to the fact that the effective network gets increasingly sparse with each iteration, and that link clustering works equally well on sparse and dense networks, whereas clique based methods requires a high level of clustering to yield any results.
We partially account for this phenomenon by differentiating between \emph{assignable} and \emph{non-assignable links}.
Links that are not part of any triangles cannot be assigned to a community by the CPA since they can never be part of a $k$-clique $(k\geq3)$.
Similarly, links that are not part of a \emph{component} that contains at least one $k$-clique $(k\geq3)$ can never be assigned by GCE, because the component contains no potential seed.

\begin{figure}
    \includegraphics[width=0.8\linewidth]{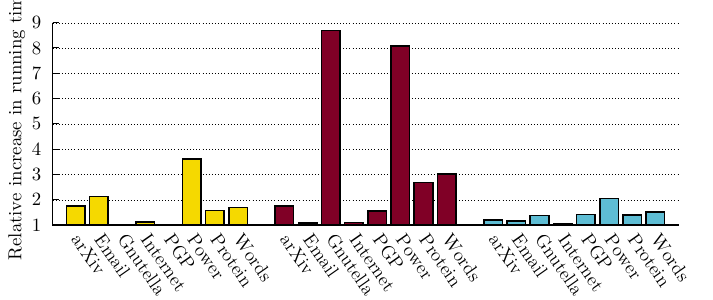}
    \caption{{\bf Average relative increase in running time due to iterated application of detection algorithms on real networks.}
    (\textit{Left}) Increase in running time for the CPA algorithm.
    (\textit{Center}) Increase in running time for the GCE algorithm.
    (\textit{Right}) Increase in running time for the LCA algorithm.
    Numerical results are summarized in the Supporting Information (S2 Table).}
    \label{Fig6}
\end{figure}

Although the increasing sparseness of the network hinders the performance of the CPA and GCE, it also reduces the cost of the subsequent detection steps.
Fig~\ref{Fig6} presents the average relative increase in running time caused by a cascading approach, for all algorithms.
Recall that the cascading detection meta-algorithm belongs to the same complexity class as the original algorithm, such that the total running time only differ by a multiplicative factor $\alpha$.
Even when accounting for the overhead associated with file handling and other miscellaneous operations, we find an average of $\alpha = 1.73$ (CPA), $\alpha=3.50$ (GCE) and $\alpha = 1.39$ (LCA)  for this set of small real networks.
These values of $\alpha$ are heavily skewed by the results on the \textit{Power} network; this network is so small ($N=4\ 941$ nodes, $L=6\ 594$ links, $L^\prime=1\ 371$ links in $k$-cliques) that the majority of the time is spent dealing with file system operations.
Gnutella is equally sparse, with less than $5\%$ of nodes belonging to at least one clique.
This leads to a similar slowdown for GCE (CPA cannot be applied more than once).

\begin{figure}
    \includegraphics[width=0.8\linewidth]{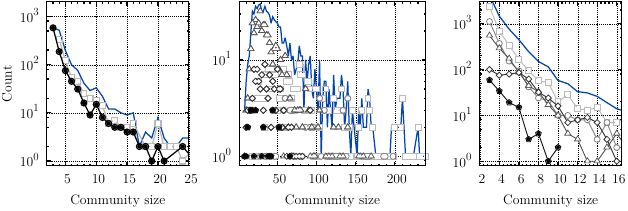}
    \caption{{\bf Distribution of the size  of the detected communities (in terms of nodes) at each iteration of the cascading approach.}
    (\emph{Left panel}) CPA,
    (\emph{Center panel}) GCE, and
    (\emph{Right panel}) LCA applied to the \textit{Words} network.
    The distributions obtained after the first iteration are shown using light gray square markers, and subsequent iterations (whenever required) are respectively marked by circles, triangles, rhombuses, pentagons and inverted triangles.
    Filled black markers indicate the last iteration.
    In the center panel, we omit for clarity the iterations that uncover very few communities (3rd, 4th, 7th and 8th).
    Interestingly, the size of the detected communities roughly follows the same distribution at each iteration.
    Therefore, the final size distribution (blue line)  has also roughly the same shape as the one obtained with standard algorithms.
    Although this is not a direct proof, it suggests that the communities unveiled through cascading are similar to the ones detected by a ``traditional'' use of the detection algorithm.
    In other words, these communities are significant and are not simple artifacts of the cascading approach.}
\label{Fig7}
\end{figure}

Fig \ref{Fig7} confirms that as the cascading detection proceeds, smaller and previously masked communities are detected, regardless of the algorithm used.
For instance, Fig \ref{Fig7} (left panel) clearly shows how a significant number of 3-cliques are overlooked by the ``traditional'' use of the CPA.
However, large communities are also found after many iterations, suggesting that the shadowing effect is not restricted to small communities.

\begin{figure}
    \includegraphics{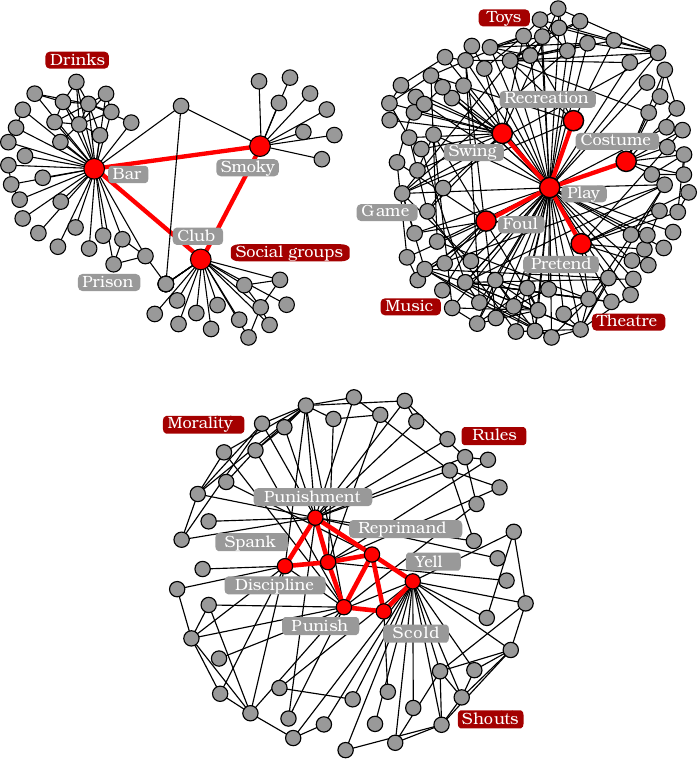}
    \caption{{\bf Sample of the communities detected with the cascading approach on the Words network.}
    (\emph{Top left panel}) Triangle detected with LCA at the third iteration.
    (\emph{Top right panel}) Star detected with LCA at the third iteration.
    (\emph{Bottom panel}) Dense community detected with the CPA at the second iteration.
    The detected communities are shown (red) as well as their neighboring nodes (grey).
    Red and grey labels identify respectively semantic fields and individual words.}
    \label{Fig8}
\end{figure}

Visual inspection of the detected communities not only verifies the quality of the hidden communities, but also confirms our intuition of the shadowing effect.
A look at Fig \ref{Fig8}~(\emph{top left panel}) shows a triangle detected at the third iteration (out of five) of the LCA on the \textit{Words} network.
This structure was missed during the initial detection due to the high degree of its three nodes, as speculated in Fig \ref{Fig4}.
Similarly, although $k=4$ was initially chosen (according to the previously discussed criterion) for the CPA on the \textit{Words} network, a second iteration using $k=3$ has permitted the detection of other significant communities such as the one shown in Fig \ref{Fig8}~(\emph{bottom panel}).

More complex structures and correlations are also brought to light using this approach.
Fig \ref{Fig8}~(\emph{top right panel}) presents a star of high-degree nodes detected at the third iteration of the LCA on the word association network.
None of these nodes are directly connected to each other, but they share many neighbors.
Hence, once the main communities were removed --- here semantic fields related to toys, theater and music --- the shadow was lifted such that this correlated, but unconnected structure, could be detected.
Whether this particular structure should be defined as a relevant community is debatable.
Keeping in mind that there are no consensus on the definition of a proper community in complex networks, the role of algorithms, and consequently of the cascading method, is to infer plausible significant structures.
%
\subsection{Real networks with meta-information}
%

Important insights can be gained by applying detection algorithms to networks that are accompanied by meta-information.
In real networks, meta-information such as declared affiliations (e.g. individuals in social networks) can be be used to define \emph{functional} groups of nodes \cite{Yang2015}.
It has been shown that, in general, functional communities do not correspond to the communities typically uncovered by detection algorithms.
In other words, the traditional mathematical definition of communities (relatively densely connected groups of nodes) seldom capture the structure of the functional subunits of real complex networks \cite{Hric2014,Yang2015}.
Regardless of this difference, identifying functional communities is often an implicit or explicit objective of detection algorithms (see for example References \cite{Ahn2010,Chen2014,Evans2009,Fortunato2010,Girvan2002,He2015,Lee2010,Newman2004,Newman2006,Xie2013}).
For comparison purposes, we therefore apply our meta-algorithm to a few real networks with functional communities inferred from meta-information.

We study 3 large networks downloaded from the SNAP database \cite{Yang2015}: the Amazon product network (\textit{Amazon}), the DBLP computer science co-authorship network (\textit{DBLP}), and the YouTube friendship social network (\textit{YouTube}).
The functional communities are determined by categories of products, conferences and journals where authors publish, and user groups, respectively.
Their properties are summarized in the Supporting Information (S3 Table).

\begin{figure}
    \includegraphics[width=0.8\linewidth]{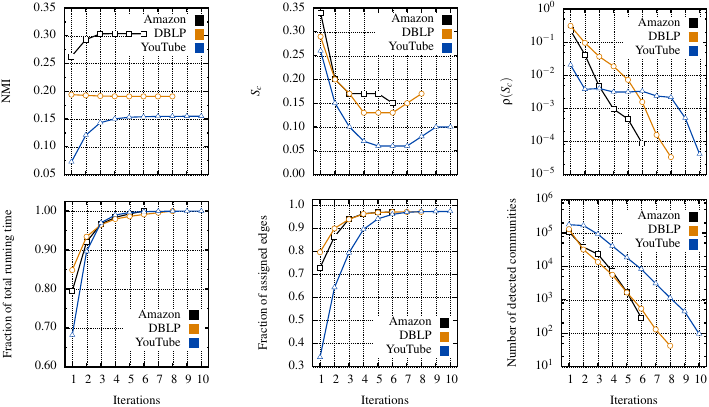}
    \caption{{\bf Case study of real networks with meta-information, using the cascading LCA meta-algorithm.}
    Complete detection is achieved with 6 iterations for \textit{Amazon} (black squares), 8 iterations for \textit{DBLP} (orange circles), and 10 iterations for the \textit{YouTube} (blue triangles).
    (\textit{Top left panel})  Normalized mutual information as a function of the number of iterations.
    (\textit{Top center panel})  Selected similarity threshold at each iteration (cf. Eq.~\eqref{Eq:Similarity_Ahn}).
    (\textit{Top right panel})  Density of the optimal link partition, in the sparser network.
    (\textit{Bottom left panel})  Elapsed fraction of the total running time, averaged over 10 
    independent realizations.
    (\textit{Bottom center panel})  Cumulative fraction of \emph{assigned} edges.
    (\textit{Bottom right panel})  Number of detected communities at each iteration.
    Numerical results are summarized in the Supporting Information (S4 Table)}
    \label{Fig9}
\end{figure}

Because the considerable time and storage space requirements of clique based algorithms make them unsuitable for such large ($L\sim10^{6}$) and clustered networks, we have restricted ourselves to the LCA algorithm for the remaining of the section.
Fig \ref{Fig9} presents the complete results of our numerical experiments.

First and most importantly, we find that the cascading approach either increases the \emph{normalized mutual information} (NMI) \cite{Lancichinetti2009b} significantly or  does not affect it (see top left panel of Fig \ref{Fig9}).
In essence, the NMI is an information theoretic measure that tells us how much information two different sets of communities share.
It penalizes over- and under-fitting, allows for overlapping communities, and is limited to the interval $[0,1]$; comparing a set of communities to itself (i.e. a perfect match) yields a NMI of 1, whereas comparing a set of communities to a random guess yields a NMI close to 0.
Hence, the increase in NMI observed for \textit{Amazon} and \textit{YouTube} means that a cascading algorithm recovers functional communities much more efficiently, whereas the quality of the description does not change significantly with subsequent iterations of the algorithm on \textit{DBLP}.

The nature of the functional communities can explain these 2 contrasting behaviors.
Indeed, the very definition of what qualifies as a link is closely related to the definition of the functional communities of the \textit{Amazon} and \textit{YouTube} networks.
Co-purchased products are likely to belong to the same category (e.g. a shirt and matching pants, a cellphone and an extra charger).
Likewise, social groups are both encouraging new friendships and emerge from existing friendships.
Hence, it should be expected that \textit{Amazon}'s and \textit{YouTube}'s functional communities are at least superficially similar to structural communities.
This intuition is confirmed by the high average density of their respective functional communities ($\langle\rho\rangle = 0.77$ and $\langle\rho\rangle=0.73$, respectively).

In contrast, the \textit{DBLP} network is a prime example of systems where links and functional communities are more loosely related.
Scientific journals and conferences are large organizations.
Many authors that never collaborated are thus regrouped in larger, sparser communities ($\langle\rho\rangle=0.52$).
However, these vast and sparse communities \emph{do not} correspond to the typical mathematical definition of a structural community.
As a result, insignificant communities are detected on average, especially since the cascading algorithm aims to uncover small and shadowed communities.
Ultimately, our analysis suggests that even if they are not ubiquitous, shadowed \emph{functional} communities naturally occur in real networks and our cascading algorithm can uncover some of them.

Second, the evolution of the structure under cascading detection is also of interest.
In all cases, the network shrinks rapidly, both in term of assignable edges (Fig \ref{Fig9}, bottom center panel) and number of detected communities (bottom right panel).
In turn, this implies that only a modest computational cost is associated to the cascading approach (bottom left panel).
The increasing sparseness of the networks (bottom center panel), resulting in a smaller similarity between links (top center panel) and a rapidly decreasing density (top right panel), suggests however that there is room for improvement.
Internal link removal is destructive since information about shadowed communities is lost in the process, as some of the internal links are shared by more than one community \cite{Ahn2010}.
Using more a more nuanced criterion for link removal could enhance the quality of the detected communities while further reducing the uncharted portion of the network \cite{Wen2011,Chen2014,He2015}.
Nevertheless, by using the simplest implementation of our cascading detection idea, we obtain surprisingly good results.
Our results suggest that shadowing is not only due to the density of the prominent communities but also to the stiffness of the resolution parameter.
In essence, by using a cascading approach, this parameter is allowed to vary artificially from a region of the network to the other, as the algorithm is effectively applied to a new network -- partially retaining the structure of the original network -- at each iteration.
A once rigid global parameter can now flexibly adapt to small changes in the topology of the network to better reveal subtle structures.

\subsection{Benchmark networks}
%
The previous sub-section indicates that the cascading approach performs better when functional communities are dense, i.e. similar to structural communities.
This hypothesis can be investigated with artificial networks generated specifically to exhibit known structural communities (called built-in communities).
We apply the cascading LCA to Lancichinetti-Fortunato \cite{Lancichinetti2009a} networks (LF networks).
LF networks can be rather precisely parametrized: in particular, one can specify the average degree $\langle k \rangle$, the mixing parameter $\mu$, the overlap fraction $f$, the membership number $\Omega$, and the scaling exponent $\tau_2$ of the community size distribution (the distribution follows a power-law).

These parameters can be used to tune the difficulty of the community detection problem.
The mixing parameter dictates the fraction of \emph{external} links.
When $\mu < 0.5$, the majority of links occur within communities, whereas links are more likely to connect different communities for $\mu>0.5$.
This latter case poses a harder challenge to detection algorithms, since communities are arguably no longer well-defined \cite{Lancichinetti2009a}.

In the standard implementation of \cite{Lancichinetti2009a}, the fraction $f$ of overlapping nodes and the membership number $\Omega$ allow for a limited control of the membership distribution (the membership $m$ of a node is the number of communities to which it belongs)
By construction, the $(1-f)N$ non-overlapping nodes have a membership of one, while the $fN$ overlapping nodes belong to $\Omega$ communities, such that the membership distribution is of the form
\begin{equation}
  p(m) = \delta_{1,m}(1-f) + \delta_{\Omega,m}f.
\end{equation}
For $f=0$, the built-in community structure consists of non-overlapping communities, whereas the system enters a highly overlapping regime when $f\to1$.
The detection problem is not necessarily more difficult at high value of $f$; it simply requires a different class of algorithms.

The last parameter of interest is $\tau_2$, the scale exponent of the community size distribution.
Since this distribution follows a power-law, its average is only well-defined for $\tau_2>2$, whereas its variance goes to infinity whenever $\tau_2<3$.
Thus, as $\tau_2$ decreases, increasingly large communities appear.
Because LF networks do not control explicitly for size correlation in neighboring communities, some of the large communities happen -- through pure chance -- to share nodes with very small communities.
In fact, it can be verified that some of these large communities neighbor much smaller communities, despite some trace of assortative mixing based on community sizes (See S2 Fig--S3 Fig).

\begin{figure}
    \includegraphics[width=0.48\linewidth]{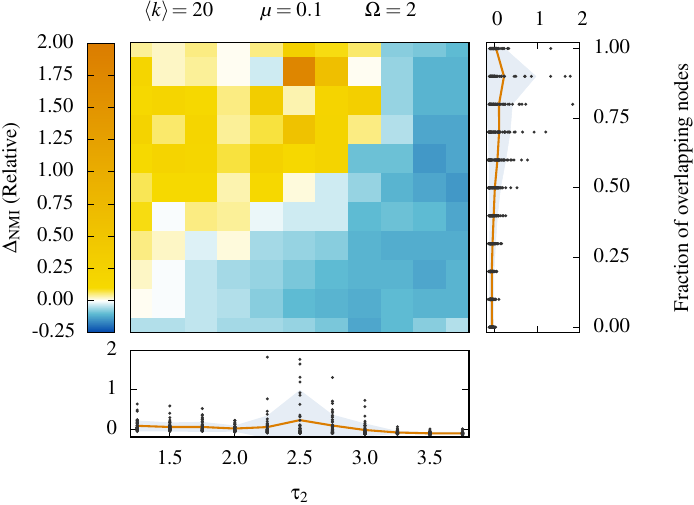}\qquad
    \includegraphics[width=0.48\linewidth]{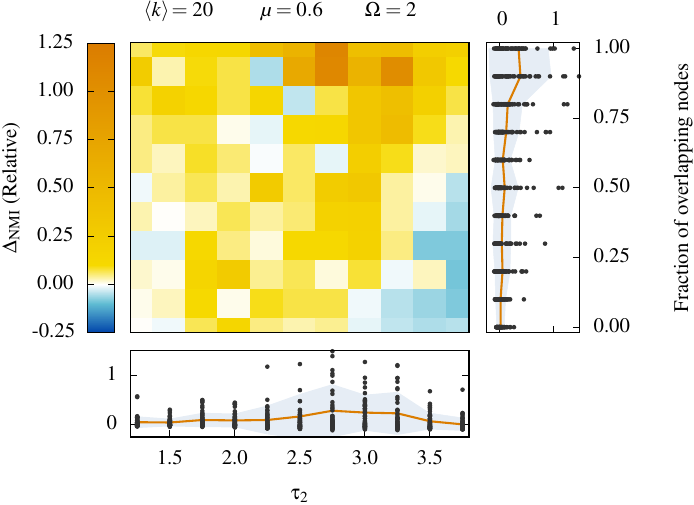}
    \caption{{\bf Comparison of the communities detected with the pure LCA and a cascading version of the LCA, for LF networks.}
    Relative change in normalized mutual information obtained by comparing the structure detected by the pure LCA and the cascading LCA, when applied to LF networks.
    All results are discrete points, but solid curves are added to guide the eye.
    (\emph{Left panel}) Lightly mixed LF networks with a mixing parameter $\mu=0.1$.
    (\emph{Right panel}) Heavily mixed LF networks $\mu=0.6$.
    We use networks of $N=5\ 000$ nodes of average degree $\langle k \rangle=20$, that belongs to $\Omega=2$ communities (\textit{if} they overlap). 
    The fraction of overlapping nodes goes from $0$ to $1$ (y-axis).
    The value of the exponent of the community size distribution $\tau_2$ ranges from 1.25 to 3.75 (x-axis).
    Averaged data is shown on the color map (10 different networks for each point), while the distribution of raw data is shown in the plots to the right and bellow (black dots).
    Within raw data plots, the solid curve shows the average and the gray area indicates the standard deviation.}
    \label{Fig10}
\end{figure}

In Fig \ref{Fig10}, we investigate a wide spectrum of possible structures by generating LF networks for $f\in[0,1]$ and $\tau_2\in[1.25,3.75]$, with $N=5\ 000$ nodes of average degree $\langle k \rangle=20$, membership number $\Omega =2$ and mixing parameters $\mu =0.1$ (left panel) and $\mu =0.6$ (right panel).
A reference NMI is computed by comparing the built-in community structure with the communities detected by the LCA (See S1 Fig).
The cascading approach is then applied to the same set of networks, and a new value of the NMI value is calculated.
The relative change in normalized mutual information $\Delta\text{NMI}$ can then be computed.

It is useful to split the joint $f\times\tau_2$ space in 4 qualitative different regions of interest to analyze the results of Fig \ref{Fig10}: 
\begin{enumerate}
  \item high $f$, low $\tau_2$: overlapping with heterogeneous community sizes;
  \item high $f$, high $\tau_2$: overlapping with homogeneous community sizes.
  \item low $f$, low $\tau_2$:  non-overlapping with heterogeneous community sizes;
  \item low $f$, high $\tau_2$: non-overlapping with homogeneous community sizes;
\end{enumerate}
Shadowing is \emph{possible} in regions where the built-in communities overlap appreciably, i.e. regions 1-2.
However, it is more \emph{likely} to occur when large community neighbors very small communities, i.e. in region 1.
This is where the cascading approach excels.
Since region 1 is where most real-world networks reside \cite{Palla2005, Ahn2010}, this further supports our claim that shadowed communities are a common occurrence in complex networks, and that cascading detection is a viable solution.

The results of regions 2,3 and 4 highlight the fact that the cascading approach is not a silver bullet.
Because shadowing happens less frequently in regions 3 and 4 (no or little overlap), repeated applications of the detection algorithm sometimes decrease the quality of the final partition, through overfitting.
Furthermore, since the effects of shadowing are more pronounced when community sizes are heterogeneous, the cascading detection approach is also prone to overfitting in regions 2 and 4.
Nonetheless, it is important to realize that the changes in NMI are \emph{relative}, and that improvements in quality are more frequent and more pronounced than their opposites (See S5 Table).

\begin{figure}
    \includegraphics[width=0.48\linewidth]{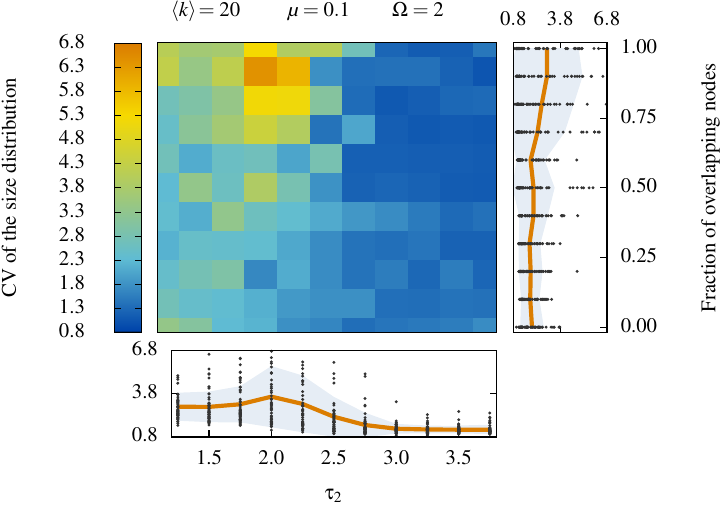}\qquad
    \includegraphics[width=0.48\linewidth]{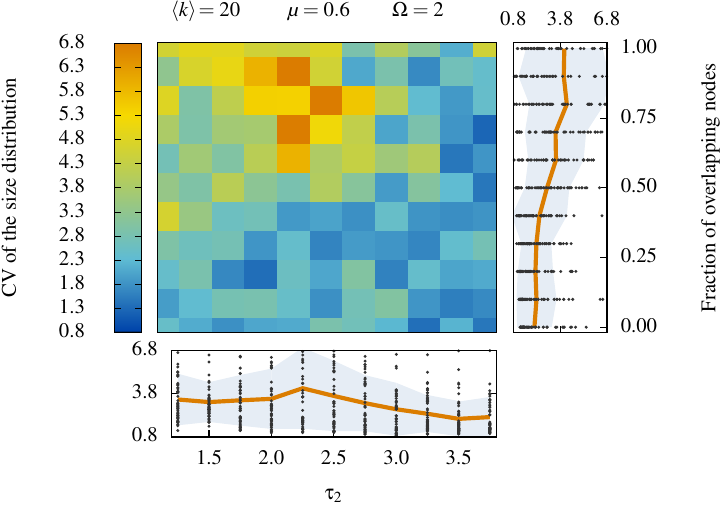}
    \caption{{\bf Heterogeneous community sizes at the initial iteration of the LCA to LF networks.}
    Coefficient of variation $c_v$ of the community size distribution for (\emph{left panel}) lightly mixed LF networks with a mixing parameter $\mu=0.1$ and (\emph{right panel}) heavily mixed LF networks $\mu=0.6$.
    The coefficient of variation is the ratio of  the standard deviation over the mean. 
    See the caption of Fig \ref{Fig10} for explanations of the layout of this figure.}
    \label{Fig11}
\end{figure}

In a normal situation, where built-in communities are not known, the results shown in Fig \ref{Fig10} call for caution, because one cannot verify the quality of the detected communities \textit{a posteriori}.
However, since we have found that the cascading approach works well on networks with overlapping communities of heterogeneous sizes, the communities detected on the first iteration can give support for the decision to cascade or not.
More precisely, if communities of homogeneous sizes are detected on the first iteration, we have no reason to expect shadowing effects to be present.
Further detection might only cause the algorithm to simply group random links together. 
Conversely, if a heterogeneous community structure is detected, than one can expect shadowing to have occurred.
In these cases, the algorithm can find remaining relevant correlations and structures to compute.
Fig \ref{Fig11} formalizes this intuition.
We confirm that the first iteration LCA captures the variability in community sizes of the built-in structure.
More importantly, we find that high heterogeneity correlates with an increase in NMI.
In general, the heterogeneity in size can be quantified by the coefficient of variation $c_v$ of the size distribution, i.e. the ratio of its standard deviation to its mean.
If we maximize the average NMI of the final outcome based on the variability of the size distribution detected at the first iteration, then $c_v \approx 1.15$ is the optimal threshold for the low mixing $(\mu=0.1)$ case, whereas the optimum threshold lies in a wide range $[0, 0.82]$ for the high mixing case.
A precise and general criterion cannot be formulated, because its specifics depends on the hidden parameters of the network and algorithm of choice, as illustrated by this example.

Nonetheless, we can assess the prevalence of shadowed \emph{structural} communities in real networks based on the above qualitative criterion, because their \emph{detected} community structure is similar to that of LF networks (See S6 Table for detailed results)
Summarily, LCA uncovers weakly mixed community structures  ($\langle \mu\rangle = 0.13$) with average memberships number ($\langle \Omega \rangle = 2.06$), and network densities ($\langle \rho \rangle = 0.002$) that all roughly correspond to the parameters used in the left panel of Fig \ref{Fig10}.
The coefficient of variations $c_v $ suggests that shadowing of structural communities definitely occurs in the \textit{Email} $(c_v = 5.51)$, and \textit{Internet} networks $(c_v = 2.42)$.
No definitive conclusion can be drawn for the other networks, since their calculated variability of $c_v>0.7$ lie slightly below the optimal threshold for low mixing cases, except for the \textit{Power} network $(c_v = 0.48)$.
These results must be considered with care since LF networks do not capture all the structural complexity of real networks \cite{Lancichinetti2009a}.
Other qualitative indicator such as the change in remaining assignable edges (Fig \ref{Fig5}), and the correlation between the size of neighboring communities pairs (See S3 Fig and analysis therein) suggest that shadowing occurs in all cases.

\begin{figure}
    \includegraphics[width=0.48\linewidth]{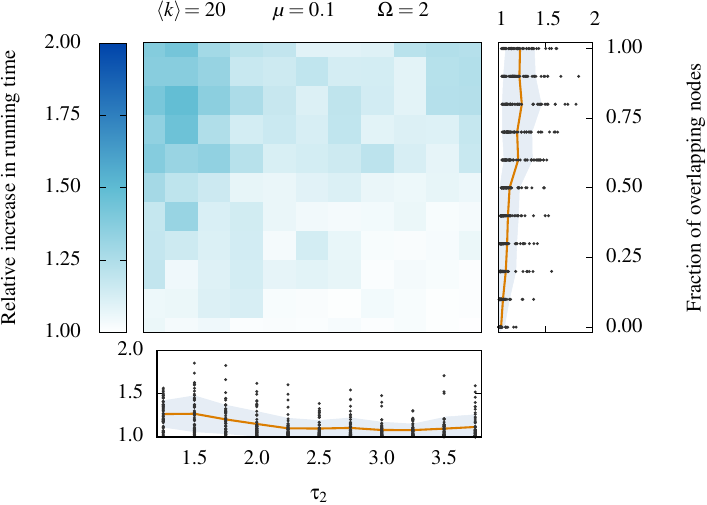}\qquad
    \includegraphics[width=0.48\linewidth]{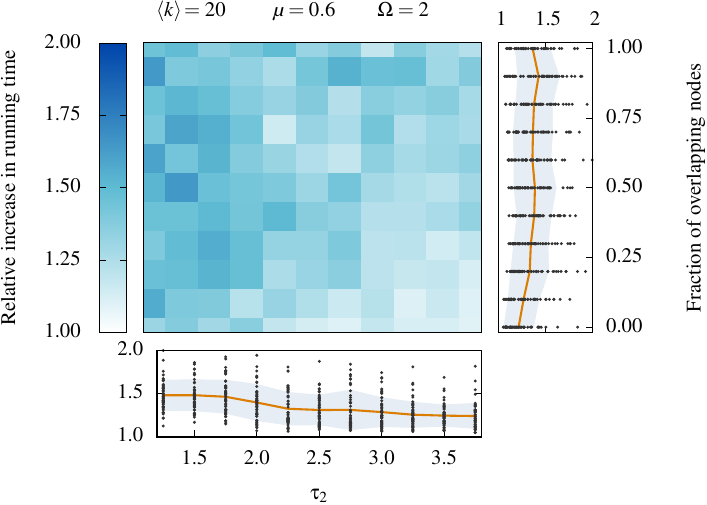}
    \caption{{\bf Relative increase in running time caused by the cascading approach.}
    Raw running times are computed to the millisecond precision and averaged over 10 independent and complete iterations.
    The network generating process is not included in the total running time.
    (\emph{Left panel}) Lightly mixed LF networks with a mixing parameter $\mu=0.1$.
    (\emph{Right panel}) Heavily mixed LF networks $\mu=0.6$.
    See the caption of Fig \ref{Fig10} for explanations of the layout of this figure.}
    \label{Fig12}
\end{figure}

Finally, we find once again that the cascading approach involves modest increase in running time (See Fig \ref{Fig12}).
Interestingly, we find that the cascading approach involves a larger increase in running time whenever the detected communities are relevant.
The Pearson correlation coefficient of the relative increase in NMI and running time amounts to 0.254 and 0.292 for $\mu = 0.1$ and  $\mu=0.6$.
The density of the remaining structures explains nicely both the increase in running time and the quality of the detected communities.
If the remaining structure is dense enough, the costly operation of detecting structural communities boosts the running time of the remaining iterations.
Conversely, if the remaining structure is sparse and uninteresting, the remaining detection steps are spent agglomerating random links together, a fast operation. 

\section{Conclusion and perspective}
In conclusion, we have defined the shadowing effect in community detection and have illustrated the types of scenarios where it might arise. 
This effect calls for a simple solution: a cascading use of detection algorithms.
This meta-approach has been shown to reduce the hidden portion of a network, and to find relevant communities in real and benchmark networks when shadowing does occur.

We have shown that a simple implementation of the cascading philosophy can indeed unveil communities that were initially overshadowed by larger and/or denser communities.
Interestingly, in both real datasets and benchmarks, cascading approach appears more likely to detect meaningful remaining communities whenever subsequent iterations of the detection algorithm still required significant computing time.
As observed in the LF benchmarks, this situation occurs with networks that we know more sensitive to shadowing effects (per definition): strongly overlapping communities with heterogeneous sizes.

The current implementation is meant to be the first level of cascading approaches, opening the way to more subtle meta-algorithms.
For instance, one could construct an extreme version, where communities are detected one by one (in the spirit of Ref. \cite{Chen2014}).
Such an approach would enable a perfect adaptation of the resolution parameter to the situation at hand.
And while it would certainly come with significant computational cost, it could lead to the mapping of the community detection problem unto simpler problems.
If we accept to detect communities one at a time, the detection of the most significant ones can be done through well optimized methods, such as modularity optimization \cite{Newman2006}, which would otherwise be incapable of detecting overlapping communities.
We could also envision cascades involving more than one algorithm to best suit the structure remaining after each iteration,
or cascades correlating the community structure obtained at one iteration with those obtained previously.
The possibilities for further developments are thus numerous, and we are hopeful that the present version may serve as a benchmark for future work. 
In fact, recent work \cite{He2015} has already expanded upon the implementation of the cascading philosophy based on an earlier pre-print version of our paper.

Any significant improvement in community detection will help shrink the gap between analytical models and their real network counterparts.
The difficult problem of accurately modeling the dynamical properties of real networks might be better tackled if one includes complex community structure through comprehensive distributions or solved motifs \cite{Allard2012,Karrer2010}, two applications for which a reliable and complete partition is fundamental.

Finally, in addition to the technical developments presented in this paper, perhaps the most insightful observation can be simply stated: since community structure occurs at all scales, global partitioning of overlapping communities must be done sequentially, cascading through the organizational layers of the network.

\section*{Acknowledgments}
The authors wish to thank the Gephi development team for their visualization tool \cite{Bastian2009}; all the authors of the cited papers for making their network data and code publicly available; and Calcul Qu\'ebec for computing facilities.


%
\newpage


\section*{Supplementary material}

# Supporting Information
# A shadowing problem in the detection of overlapping communities: lifting the resolution limit through a cascading procedure

Jean-Gabriel Young,[1] Laurent Hébert-Dufresne,[1,2] Antoine Allard,[1,3] and Louis J. Dubé[1]

[1]*Département de Physique, de Génie Physique, et d'Optique,*
*Université Laval, Québec (Québec), Canada, G1V 0A6*
[2]*Santa Fe Institute, Santa Fe, NM 87501, USA*
[3]*Departament de Féisica Fonamental, Universitat de Barcelona, Barcelona, Espanya, 08028*

Table S1: Description and properties of real networks used in this study.

| Network | $N^a$ | $L^b$ | $L'^c$ | $L''^d$ | $\langle k \rangle^e$ | $C^f$ | Reference |
|---|---|---|---|---|---|---|---|
| arXiv | 30 561 | 125 959 | 124 930 | 121 006 | 8.24 | 0.63 | [Palla 2005] |
| Email | 1 134 | 5 143 | 5 143 | 5 055 | 9.07 | 0.46 | [Guimera 2003] |
| Gnutella | 36 682 | 88 328 | 88 303 | 4 056 | 4.82 | 0.01 | [Ripeanu 2002] |
| Internet | 22 963 | 48 436 | 48 436 | 24 170 | 4.22 | 0.23 | [Hébert-Dufresne 2011] |
| PGP | 10 680 | 24 316 | 24 316 | 17 135 | 4.55 | 0.27 | [Boguñá 2004] |
| Power | 4 941 | 6 594 | 6 594 | 1 371 | 2.67 | 0.08 | [Watts 1998] |
| Protein | 2 640 | 6 379 | 6 296 | 3 721 | 4.83 | 0.21 | [Palla 2005] |
| Words | 7 207 | 31 784 | 31 784 | 22 169 | 8.82 | 0.15 | [Palla 2005] |

[a] Number of nodes.
[b] Number of links. We use an undirected projection where all multi-edges are only counted once, and where self-loops are removed.
[c] Number of *assignable* links for the GCE algorithm, i.e. links that are part of a component which contains at least one 3-clique.
[d] Number of *assignable* links for the CPA, i.e. links that are part of at least one 4-clique.
[e] Average degree, with $\langle k \rangle = 2L/N$.
[f] Average clustering coefficient, i.e. average fraction of existing edges between neighbors of a node.

Table S2: Summary of the results presented in Fig. 5. and Fig. 6.

|  | arXiv | Email | Gnutella | Internet | PGP | Power | Protein | Words |
|---|---|---|---|---|---|---|---|---|
| $\text{CPA}_s{}^a$ | 54.3 | 74.8 | 0 | 95.9 | 47.0 | 79.2 | 45.9 | 50.6 |
| $\text{CPA}_f{}^b$ | 7.7 | 65.8 | 0 | 88.0 | 6.7 | 5.5 | 18.1 | 22.0 |
| Iterations$^c$ | 3 | 2 | 1 | 4 | 2 | 2 | 2 | 2 |
| $\Delta T^d$ | 1.76 | 2.12 | 1.00 | 1.11 | 1.00 | 3.60 | 1.58 | 1.70 |
| $\text{GCE}_s{}^a$ | 26.9 | 64.9 | 99.7 | 40.1 | 32.4 | 92.4 | 51.7 | 54.5 |
| $\text{GCE}_f{}^b$ | 2.9 | 12.1 | 84.9 | 9.6 | 12.5 | 63.0 | 21.6 | 14.9 |
| Iterations$^c$ | 8 | 6 | 4 | 6 | 5 | 4 | 6 | 8 |
| $\Delta T^d$ | 1.76 | 1.08 | 8.69 | 1.11 | 1.56 | 8.08 | 2.69 | 3.02 |
| $\text{LCA}_s{}^a$ | 22.3 | 47.7 | 35.3 | 28.2 | 46.3 | 63.6 | 38.5 | 46.1 |
| $\text{LCA}_f{}^b$ | 2.3 | 1.3 | 18.7 | 2.7 | 4.5 | 4.4 | 6.2 | 2.6 |
| Iterations$^c$ | 7 | 5 | 2 | 5 | 6 | 3 | 3 | 5 |
| $\Delta T^d$ | 1.20 | 1.16 | 1.38 | 1.05 | 1.41 | 2.05 | 1.39 | 1.51 |

[a] Percentage of remaining assignable links for a *standard* use of the algorithm.
[b] Percentage of remaining assignable links after the cascading approach is applied.
[c] Number of applications of the cascading algorithm before the final state is reached.
[d] Relative increase in running time of the complete algorithm, averaged over 10 independent realizations, timed at the millisecond precision.

Table S3: Description and properties of real networks used in this study.

| Network | $N^a$ | $L^b$ | $\langle k \rangle^c$ | $C^d$ | $g^e$ | $\langle s \rangle^f$ | $\langle \rho \rangle^g$ | Reference |
|---|---|---|---|---|---|---|---|---|
| Amazon | 334 863 | 925 872 | 5.53 | 0.40 | 271 570 | 11.7 | 0.77 | [Yang 2015] |
| DBLP | 317 080 | 1 049 866 | 6.62 | 0.63 | 13 477 | 53.4 | 0.52 | [Yang 2015] |
| YouTube | 1 134 890 | 2 987 624 | 5.27 | 0.08 | 16 386 | 7.88 | 0.73 | [Yang 2015] |

[a] Number of nodes.
[b] Number of links. An undirected projection without multi-edges and self-loops is used.
[c] Average degree, with $\langle k \rangle = 2L/N$.
[d] Average clustering coefficient, i.e. average fraction of existing edges between neighbors of a node.
[e] Number of structural communities
[f] Average size, in nodes, of functional communities
[g] Average density of functional communities

Table S4: Summary of the results presented in Fig. 9.

| Network | LCA$_s$[a] | LCA$_f$[b] | It.[c] | $\Delta T$[d] | $\Delta$NMI[e] |
|---|---|---|---|---|---|
| Amazon | 27.1 | 2.9 | 6 | 1.28 | 0.156 |
| DBLP | 20.3 | 2.8 | 8 | 1.19 | -0.016 |
| YouTube | 66.0 | 2.6 | 10 | 1.47 | 1.128 |

[a] Percentage of remaining assignable links for a *standard* use of the algorithm.
[b] Percentage of remaining assignable links after the cascading approach is applied.
[c] Number of applications of the cascading algorithm before the final state is reached.
[d] Relative increase in running time of the complete algorithm, averaged over 10 independent realizations, timed at the millisecond precision.
[e] Relative change in normalized mutual information.

Table S5: Summary of the results presented in Fig. 10.

| $\mu$[a] | $\langle\{\Delta\text{NMI}\}\rangle$[b] | $\sigma(\{\Delta\text{NMI}\})$[c] | $\min(\{\Delta\text{NMI}\})$[d] | $\max(\{\Delta\text{NMI}\})$[e] |
|---|---|---|---|---|
| 0.1 | 0.023 | 0.29 | -0.17 | 4.79 |
| 0.6 | 0.018 | 0.31 | -0.11 | 2.25 |

[a] Mixing parameter of the LF network.
[b] Average relative difference in NMI.
[c] Standard deviation of the distribution of relative difference in NMI.
[d] Minimum of the distribution of relative difference in NMI.
[e] Maximum of the distribution of relative difference in NMI.

Table S6: Properties of the detected community structure of real networks.

| | | arXiv | Email | Gnutella | Internet | PGP | Power | Protein | Words |
|---|---|---|---|---|---|---|---|---|---|
| CPA | $\hat{\mu}$[a] | 0.05 | 0.12 | 0.06 | 0.00 | 0.02 | 0.01 | 0.07 | 0.22 |
| | $c_v$[b] | 0.56 | 1.11 | 0.43 | 0.73 | 1.16 | 0.33 | 0.83 | 1.27 |
| | $\langle\Omega\rangle$[c] | 1.69 | 1.47 | 1.13 | 1.24 | 1.15 | 1.15 | 1.15 | 1.49 |
| GCE | $\hat{\mu}$[a] | 0.11 | 0.11 | 0.00 | 0.07 | 0.03 | 0.01 | 0.09 | 0.24 |
| | $c_v$[b] | 1.03 | 0.15 | 0.44 | 1.85 | 1.25 | 0.51 | 0.94 | 0.72 |
| | $\langle\Omega\rangle$[c] | 1.54 | 1.82 | 1.02 | 1.22 | 1.20 | 1.12 | 1.09 | 1.59 |
| LCA | $\hat{\mu}$[a] | 0.08 | 0.23 | 0.12 | 0.05 | 0.08 | 0.08 | 0.12 | 0.25 |
| | $c_v$[b] | 0.74 | 2.42 | 0.89 | 5.51 | 0.78 | 0.48 | 0.70 | 0.91 |
| | $\langle\Omega\rangle$[c] | 2.96 | 2.85 | 1.69 | 1.94 | 2.13 | 1.28 | 1.80 | 2.84 |

[a] Estimated mixing parameter.
[b] Coefficient of variation of the size distribution.
[c] Average membership number.

The mixing parameter is given by the average of the ratio $1 - k_i^{(in)}/k_i$ taken over all nodes $i$, where $k_i^{(in)}$ is the number of neighbors of node $i$ which share at least a community with $i$, and where $k_i$ is the degree of node $i$ [Lancichinetti et al., PRE 80, 2009]. The coefficient of variation $c_v$ is defined as the standard deviation of a distribution, normalized by the mean.

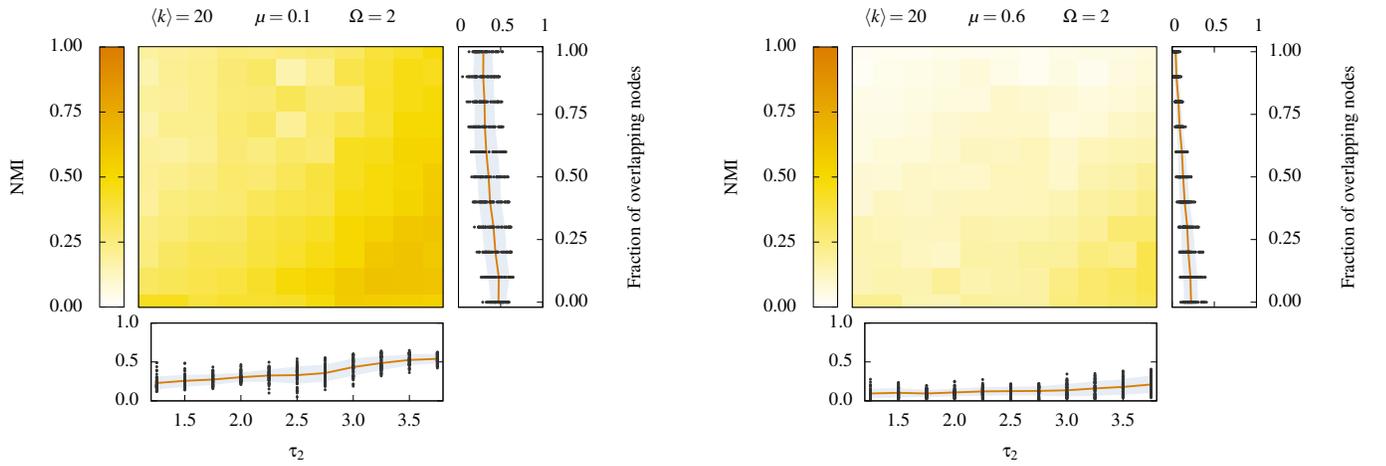

S1 FIG. **Absolute value of the normalized mutual influence for a normal use of LCA on LF networks.** This figure shows the average value of the NMI *before* any further detection steps are performed, i.e. it illustrates the results of the pure LCA. (*Left panel*) Lightly mixed LF networks with a mixing parameter $\mu = 0.1$. (*Right panel*) Heavily mixed LF networks $\mu = 0.6$. We use networks of $N = 5\,000$ nodes of average degree $\langle k \rangle = 20$, that belongs to $\Omega = 2$ communities (*if* they overlap).

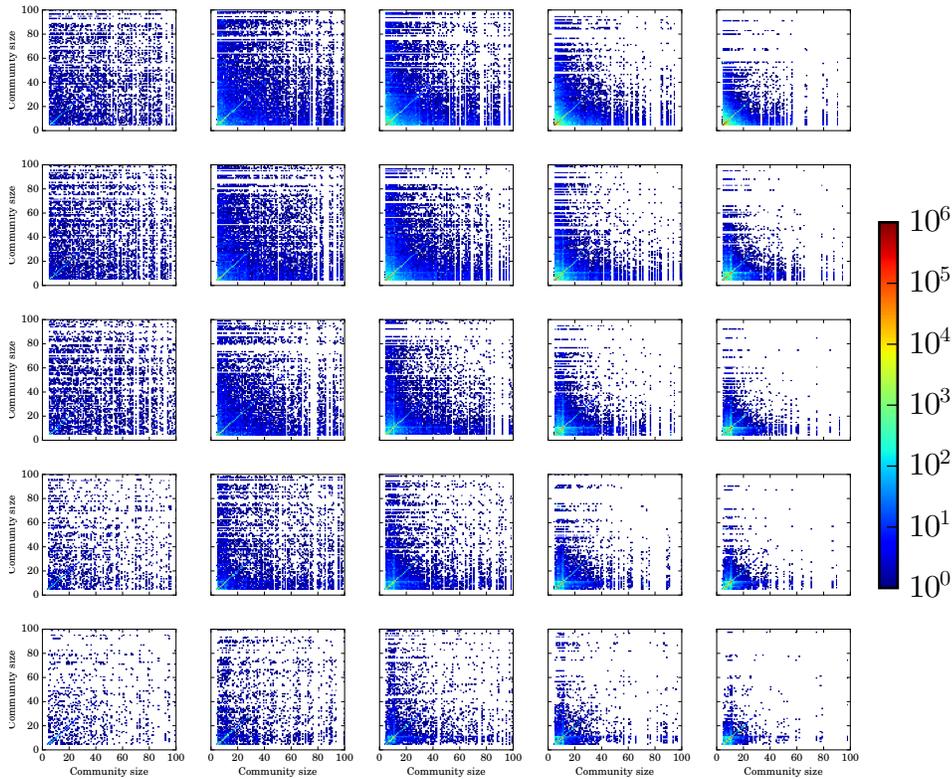

S2 FIG. **Correlation between the size of neighboring built-in communities in LF benchmark networks.** This array samples 25 points of the parameter space, i.e. fractions of overlapping nodes set to $f = [1/5, 2/5, 3/5, 4/5, 1]$ (bottom to top) and size distribution exponents set to $\tau_2 = [1.5, 2, 2.5, 3, 3.5]$ (left to right). All networks consists of $N = 5\,000$ nodes of average degree $\langle k \rangle = 20$, that belongs to $\Omega = 2$ communities (*if* they overlap), with a mixing parameter $\mu = 0.1$. Each subplot shows the number of communities of $x$ nodes in the neighborhood of communities of $y$ nodes, for all $x, y \leq 100$. Communities that share at least one node are defined as neighbors. In all cases, large communities are found in the vicinity of smaller communities of all sizes (off-diagonal elements are present). This pattern is however more pronounced in the highly heterogeneous region (right-hand side of the figure), where shadowing occurs (See Fig 10 of main text). The correlation patterns are essentially the same in the high mixing case $\mu = 0.6$ (not shown).

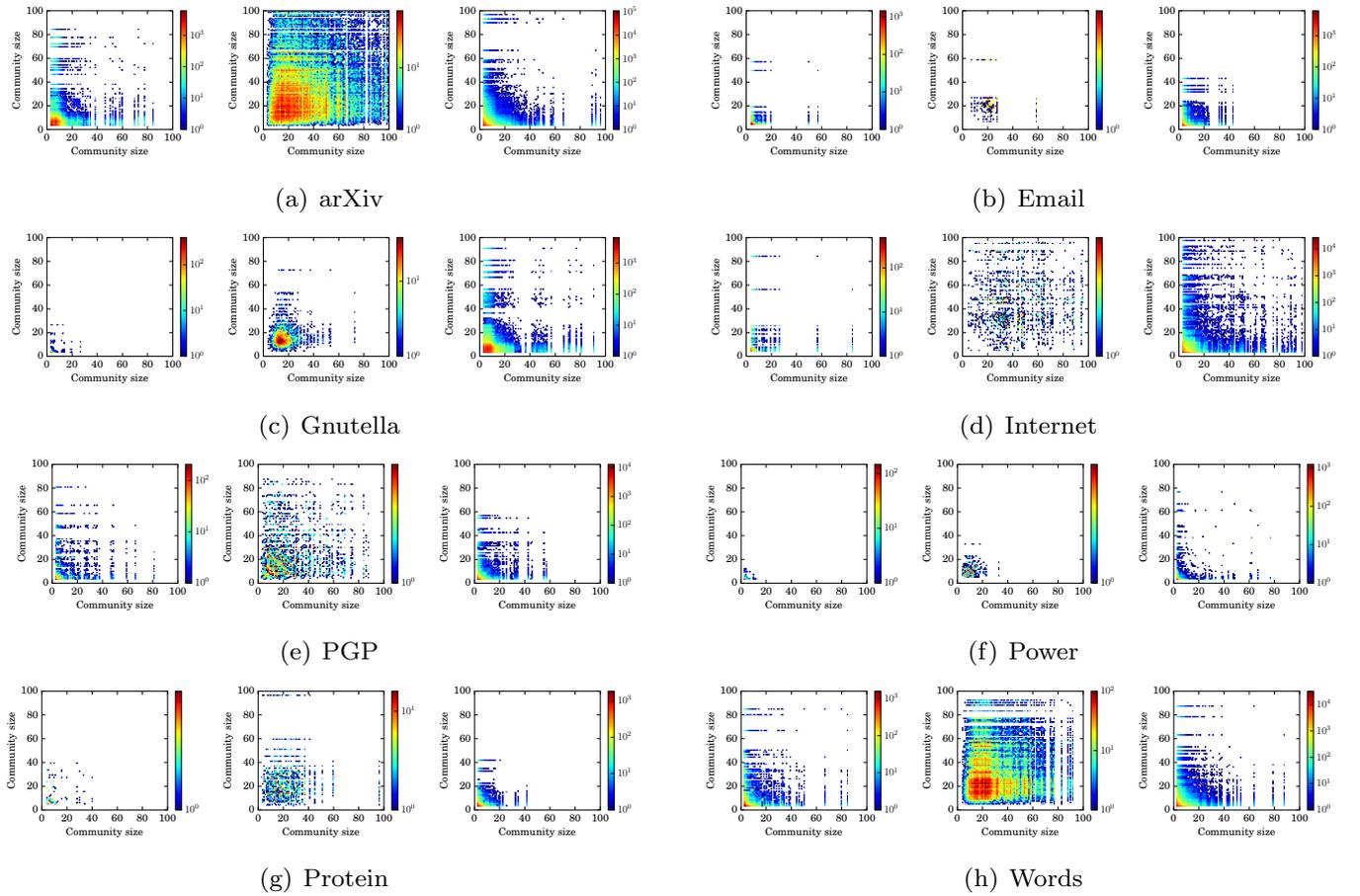

S3 FIG. **Correlation between the size of neighboring communities in real networks** For each network, correlations are computed using the community structure detected by the cascading version of CPA (left), GCE (center) and LCA (right). Each subplot shows the number of communities of $x$ nodes in the neighborhood of communities of $y$ nodes, for all $x, y \leq 100$. Communities that share at least one node are defined as neighbors. To identify the possibility of shadowed communities, one must look for large communities in the vicinity of small ones, i.e. for off diagonal elements in the rows / columns corresponding to community sizes that are present in the network. For CPA, one need not look far away from the diagonal, since small communities can shadow even smaller communities (see Fig 1 of the main text). For the other two algorithms, larger differences in sizes are required (see Figs 2 and 4 of the main text). These observations suggests that shadowing occurs in all cases (well populated correlation diagrams).



\end{document}